\documentclass[12pt]{article}
\usepackage{latexsym,amsmath}
\usepackage[dvips]{graphicx}
\usepackage{lscape}
\usepackage{color}
\usepackage[pagewise]{lineno}

\usepackage{graphicx} 
\usepackage {epstopdf}
\usepackage {subfigure}
\usepackage{hyperref}
\usepackage{booktabs} 
\usepackage{amsmath}
\usepackage{multicol}
\usepackage{multirow}
\usepackage{xcolor}
\usepackage{float}
\usepackage{amsthm}
\usepackage{amsfonts}
\usepackage[margin=1in]{geometry}
\usepackage[toc]{appendix}
\usepackage{stackengine}
\usepackage{bbold}
\RequirePackage[OT1]{fontenc}
\usepackage[authoryear]{natbib}
\usepackage{adjustbox}
\usepackage{amsmath,amssymb,amsthm,bm,latexsym}
\usepackage{graphicx,psfrag,epsf}
\usepackage[ruled,vlined]{algorithm2e}
\usepackage{epigraph,xcolor}
\usepackage{enumerate}
\usepackage{url}
\usepackage{booktabs}
\usepackage{multirow}
\usepackage{hyperref}
\usepackage{pifont}
\hypersetup{colorlinks=true,citecolor=blue,
	pdfpagemode=FullScreen,linktocpage=true}
\usepackage{cleveref}
\usepackage{amsfonts}
\usepackage{amsmath}
\usepackage{orcidlink}
\usepackage{amsthm}
\usepackage[T1]{fontenc}
\usepackage{multirow}
\usepackage{array}
\usepackage{booktabs}

\newtheorem{theorem}{Theorem}

\def\doublespace{\baselineskip=24pt}

\newcommand{\uiota}             {\mbox{\boldmath$\uiota$}}

\def\beq{\begin{equation}}
\def\eeq{\end{equation}}
\def\beqa{\begin{eqnarray}}
\def\eeqa{\end{eqnarray}}
\def\beqan{\begin{eqnarray*}}
\def\eeqan{\end{eqnarray*}}

\def\bc{\begin{center}}
\def\ec{\end{center}}
\def\btable{\begin{table}[htbp]}
\def\etable{\end{table}}
\def\bfig{\begin{figure}[htbp]}
\def\efig{\end{figure}}

\def\bi{\begin{itemize}}
\def\ei{\end{itemize}}
\def\R{\mathbb{R}}

\newcommand{\convD}{\xrightarrow[]{d}}
\newcommand{\convP}{\xrightarrow[]{P}}

\newcommand{\RNum}[1]{\uppercase\expandafter{\romannumeral #1\relax}}

\newcommand{\norm}[1]{\left\| #1 \right\|}

\numberwithin{equation}{section}  
\newtheoremstyle{general}
{3mm} 
{3mm} 
{\it} 
{} 
{\bfseries} 
{.} 
{.5em} 
{} 

\theoremstyle{general}

\newtheoremstyle{bolddef} 
  {\topsep}              
  {\topsep}              
  {\normalfont}          
  {}                     
  {\bfseries}            
  {.}                    
  { }                    
  {\thmname{#1}\thmnumber{ #2}\thmnote{ (#3)}} 

\theoremstyle{bolddef}

\newtheorem{assumption}{Assumption}

\begin{document}

\doublespace
\begin{center}
    \noindent{\huge\bf  Testing Equality of Medians for Multiple Samples
 }
\end{center}

\begin{center}
    \noindent Swapnaneel Bhattacharyya  \\
\
\noindent{ Indian Statistical Institute, Kolkata}\\
\
\noindent{
\href{mailto:swapnaneelbhattacharyya@gmail.com}{swapnaneelbhattacharyya@gmail.com}}
\end{center}

\noindent{\bf Abstract}: In this paper, we construct a consistent non-parametric test for testing the equality of population medians for different samples when the observations in each sample are independent and identically distributed. This test can be further used to test the equality of unknown location parameters for different samples. The method discussed in this paper can be extended to any quantile level instead of the median. We present the theoretical results and also demonstrate the performance of this test through simulation studies.  



\tableofcontents

\newpage

\section{Introduction}
Testing the equality of Medians for different populations is one of the very famous problems in statistics. For example, in a medical study comparing a new drug with a standard treatment, the response from the majority of patients can be more significant than the average response (\cite{cox1985testing}). Similar situations arise in various fields, such as biomedical research (\cite{albers1984approximate}), medical diagnosis (\cite{rudolfer1985large}), and the wood industry (\cite{huang2001confidence}). 

\cite{marshall1950some} were the first to consider comparing the percentage points of two arbitrary continuous populations and proposed a nonparametric test, which was later extended and modified by \cite{walsh1954bounded}. \cite{albers1984approximate} addressed a biomedical problem and provided an approximate distribution-free confidence interval for the difference between two quantiles, which was further refined by \cite{bristol1990distribution}. \cite{wilcox1991testing} and \cite{wilcox1995comparing} proposed a method to compare two groups based on multiple quantiles. For comparing the quantiles of two normal populations, Campbell and Rudolfer (1981) suggested a large-sample test utilizing the Welch test statistic used in the Behrens–Fisher problem, a method further discussed by \cite{rudolfer1985large}. Additionally, \cite{cox1985testing} explored the equality of two normal percentiles through the generalized likelihood ratio test, an adjusted version of this test, and a test statistic based on Cochran’s test. \cite{johnson2003some} proposes asymptotic methods for inference about the ratio of two percentiles. \cite{hin2012rank} comes up with a rank-based method motivated by the Kruskal-Wallis test. 

Most of the above tests are either valid for two samples, need distributional assumptions, or lack consistency. In this paper, We propose an asymptotically distribution-free consistent test to assess the equality of population medians across different samples, assuming that the observations within each sample are independent and identically distributed. This test can also be applied to test the equality of unknown location parameters across different samples. Furthermore, the approach discussed in this paper can be extended to any quantile level, not just the median. The paper is organized as follows: In \Cref{sec: results}, we provide the theoretical discussions along with the test statistic and its properties. The proofs of the theorems presented in this section can be found in the appendix. In \Cref{sec: simu}, we demonstrate the power of our proposed test across the median differences to access the specificity of the test. 

\section{Results and Discussion}
\label{sec: results}

Let 
\[
\begin{aligned}
    X_{1,1}, \ldots, X_{1,n_1} & \overset{\text{iid}}{\sim} F_1, \\
    X_{2,1}, \ldots, X_{2,n_2} & \overset{\text{iid}}{\sim} F_2, \\
    & \vdots \\
    X_{k,1}, \ldots, X_{k,n_k} & \overset{\text{iid}}{\sim} F_K
\end{aligned}
\]
be $k-$many independent samples. Let $M_i$ be the unique median of $F_i$. Let $\widehat{M}_{n_i, i}$ be the sample median of the $i$-th sample, $1 \leq i \leq k$. We write $\widehat{M}_{n_i, i} \equiv \widehat{M}_i$. We have to test the hypothesis:
\begin{equation}
\label{eq: hypo}
    \begin{split}
        H_0 &: M_1 = \cdots = M_k \\
        H_1 &: H_0 \text{ is false.}
    \end{split}
\end{equation}
We start with the following assumptions. 
\begin{assumption}[Assumption on Sample Sizes]
\label{ass: sample size}
$\lim_{n_i \to \infty} \frac{n_i}{n_1} = \lambda_i, \quad 1 \leq i \leq k$, where $\lambda_i \in (0,\infty)$. 
\end{assumption}

\begin{assumption}[Assumption of the distributions]

(2.1) $F_i$ has a density $f_i$ satisfying the Lipschitz condition of order 1 i.e., $|f_i(x) - f_i(y)| \leq c|x-y|, \quad \forall x, y \in \mathbb{R}$ \\
(2.2) $F_i$ has unique Median $M_i$ satisfying $f_i(M_i) > 0$. \\
(2.3) $f_i$ is differentiable at the neighborhood of $M_i$, and $f_i'(x)$ is bounded in that neighborhood. \\
(2.4) The characteristic function (CHF) $\phi_i(\cdot)$ of $X_{i,1}$ satisfies:
\[
\int_{-\infty}^\infty |\phi_i(t)| \, dt < \infty.
\]
\end{assumption}
For a sample consisting of $n$ many observations $X_1,\cdots,X_n$ and the bandwidth $b_n$, define the averaging kernel by
\[
K_n(x) = \frac{1}{b_n} K\left(\frac{x}{b_n}\right),
\]
and let the corresponding kernel density estimator be
\[
\widehat{f(x)} = \frac{1}{n} \sum_{i=1}^n K_n(x - X_i).
\]
In our case, for the $i-$th sample, based on the $n_i$ observations $\{X_{i,1},\cdots,X_{i,n_i}\}$, let the kernel density estimator be $\widehat{f}_i(x) : 1 \leqslant i \leqslant k$. We choose the kernel function $K$ and the bandwidth $b_n$ satisfying the following assumptions. 

\begin{assumption}[Assumption on the Kernel Function]
The kernel function $K$ is a density function satisfying a Lipschitz condition of order 1, that is,
\[
|K(x) - K(y)| \leq c|x - y|,
\]
and
\[
\int_{-\infty}^\infty |x| |K(x)| \, dx < \infty.
\]
\end{assumption}
\begin{assumption}[Assumption on the Bandwidth]
The bandwidth $b_n$ tends to zero slowly enough such that:
\[
n b_n (\log n)^{-1} \to +\infty,
\]
\[
\sum_{n=1}^\infty (b_n)^{-9/2} (\log n)^{-3/2} n^{-5/2} < \infty,
\]
\[
n^{1/2} b_n^{3/2} (\log n)^{-1/2} \to 0.
\]
\end{assumption}
Then, under assumption (3) and (4), from Theorem 3.1 of \cite{tran1992kernel}, it follows that for any compact set $D \subset \R$, 
\[
\sup_{x \in D} |\widehat{f_i(x)} - f_i(x)| = O(\Psi(n_i)) \quad \text{a.s.},
\] where $\Psi(n) = \left( \log n \cdot (n b_n)^{-1} \right)^{1/2}, 1 \leqslant i \leqslant k$. Then, from \cite{bahadur1966note}, under Assumptions (2.2) and (2.3), the sample median of the $i-$th sample ($\widehat{M_i}$) can be written as, 
\[
\widehat{M}_i = M_i + \frac{\frac{1}{2} - \widehat{F}_{n_i,i}(M_i)}{f_i(M_i)} + R_{n_i}
\]
where with probability 1, 
\[
R_n = O\left(n^{-3/4} (\log n)^{1/2} (\log \log n)^{1/4}\right).
\]
and $\widehat{F}_{n_i,i}$ is the empirical CDF of the $i-$th sample. Furthermore, under the assumption that $f_i$ is continuous (which follows from Assumption (2.1)) and Assumption (2.2), the sample medians satisfy the asymptotic normality, i.e. as $n_i \to \infty$,
\[
\sqrt{n_i} (\widehat{M_i} - M_i) \convD N\left(0, \frac{1}{4 f_i(M_i)^2}\right).
\]
So when Assumptions (1)-(4) are satisfied, using the sample medians we can construct a test for testing the equality of population medians. 
Let $\bm{\Sigma}$ be the $k \times k$ matrix:
\begin{equation}
\label{eq: Sigma}
    \bm{\Sigma} = \text{diag} \left( \frac{1}{4 \lambda_1 f_1(M_1)^2}, \ldots, \frac{1}{4 \lambda_k f_k(M_k)^2} \right)
\end{equation}
and $\widehat{\bm{\Sigma}}$ be its estimate defined as:
\begin{equation}
    \label{eq: Sigma Hat}
    \widehat{\bm{\Sigma}} = \text{diag} \left( \frac{1}{4 \lambda_1 \widehat{f}_1(\widehat{M_1})^2}, \ldots, \frac{1}{4 \lambda_k \widehat{f}_k(\widehat{M_k})^2} \right).
\end{equation}
and let $A$ be the $(k-1) \times k$ matrix:
\begin{equation}
\label{eq: A}
    A = 
\begin{bmatrix}
-1 & 1 & 0 & \cdots & 0 & 0 \\
0 & -1 & 1 & \cdots & 0 & 0 \\
\vdots & \vdots & \vdots & \ddots & \vdots & \vdots \\
0 & 0 & 0 & \cdots & -1 & 1
\end{bmatrix}_{(k-1) \times k}.
\end{equation}

\begin{theorem}
\label{th: th1}
If Assumptions (1),(2.1),(2.2),(2.4),(3) and (4) are satisfied, then:
\[
T_{n_1} := n_1 \begin{bmatrix}
\widehat{M_1} - \widehat{M_2} - (M_1 - M_2) \\
\widehat{M}_2 - \widehat{M}_3 - (M_2 - M_3) \\
\vdots \\
\widehat{M_{k-1}} - \widehat{M_k} - (M_{k-1} - M_k)
\end{bmatrix}^T
\left( A \widehat{\bm{\Sigma}} A^T \right)^{-1}
\begin{bmatrix}
\widehat{M_1} - \widehat{M_2} - (M_1 - M_2) \\
\widehat{M}_2 - \widehat{M}_3 - (M_2 - M_3) \\
\vdots \\
\widehat{M_{k-1}} - \widehat{M_k} - (M_{k-1} - M_k)
\end{bmatrix}
\xrightarrow{d} \chi^2_{k-1} 
\]
as  $n_1, \ldots, n_k \to \infty$.
 \end{theorem}

So in view of \Cref{th: th1}, we can now define our test statistic to be 
\begin{equation}
    T_{n_1,H_0} := n_1 \begin{bmatrix}
\widehat{M_1} - \widehat{M_2} \\
\widehat{M}_2 - \widehat{M}_3  \\
\vdots \\
\widehat{M_{k-1}} - \widehat{M_k}
\end{bmatrix}^T
\left( A \widehat{\bm{\Sigma}} A^T \right)^{-1}
\begin{bmatrix}
\widehat{M_1} - \widehat{M_2}  \\
\widehat{M}_2 - \widehat{M}_3  \\
\vdots \\
\widehat{M_{k-1}} - \widehat{M_k} 
\end{bmatrix}
\end{equation}
\textbf{Remark 1:} The test 
\begin{equation}
    \phi := \mathbb{1}\left( T_{n_1,H_0} > \chi^2_{k-1,1-\alpha} \right)
\end{equation}
is an asymptotically level $\alpha$, consistent test for testing the hypothesis (\ref{eq: hypo}). 

To understand the rate of consistency of the above test, we obtain the following approximation theorem which establishes a high probability upper bound on our test statistic. 

\begin{theorem}
    Let $M := 
\begin{bmatrix}
M_1 - M_2 \\
\vdots \\
M_{k-1} - M_k
\end{bmatrix}, \quad
\widehat{M} := 
\begin{bmatrix}
\widehat{M}_1 - \widehat{M}_2 \\
\vdots \\
\widehat{M}_{k-1} - \widehat{M}_k
\end{bmatrix}.$ If Assumptions (1)-(4) are satisfied then for some $c > 0$, 
\begin{align*}
    &\hspace{0.5cm}\left| \widehat{M}^T \left(A \widehat{\Sigma} A^T\right)^{-1} \widehat{M} - M^T \left(A \Sigma A^T\right)^{-1} M \right| \\
    &\leqslant c \left[ R_n + \sum_{i=1}^{k-1} \left|  \frac{\frac{1}{2} - \widehat{F}_{n_i,i}(M_i)}{f_i(M_i)} -  \frac{\frac{1}{2} - \widehat{F}_{n_{i+1},i+1}(M_{i+1})}{f_{i+1}(M_{i+1})} \right| + \|\widehat{\Sigma}^{-1} - \Sigma^{-1}\|_F \right]
\end{align*}
with probability $1 - o(1)$ as $n_1,\cdots,n_k \to \infty$ where $R_n = O\left(n^{-3/4} (\log n)^{1/2} (\log \log n)^{1/4}\right)$ with $n = \max\{n_i\}_{i=1}^k$ and $\norm{\cdot}_F$ denotes the Frobenius Norm of Matrices.
\end{theorem}

\section{Simulation Studies}
\label{sec: simu}
\vspace{-0.5cm}
\begin{figure}[H]
    \centering
    \includegraphics[width=0.6\textwidth]{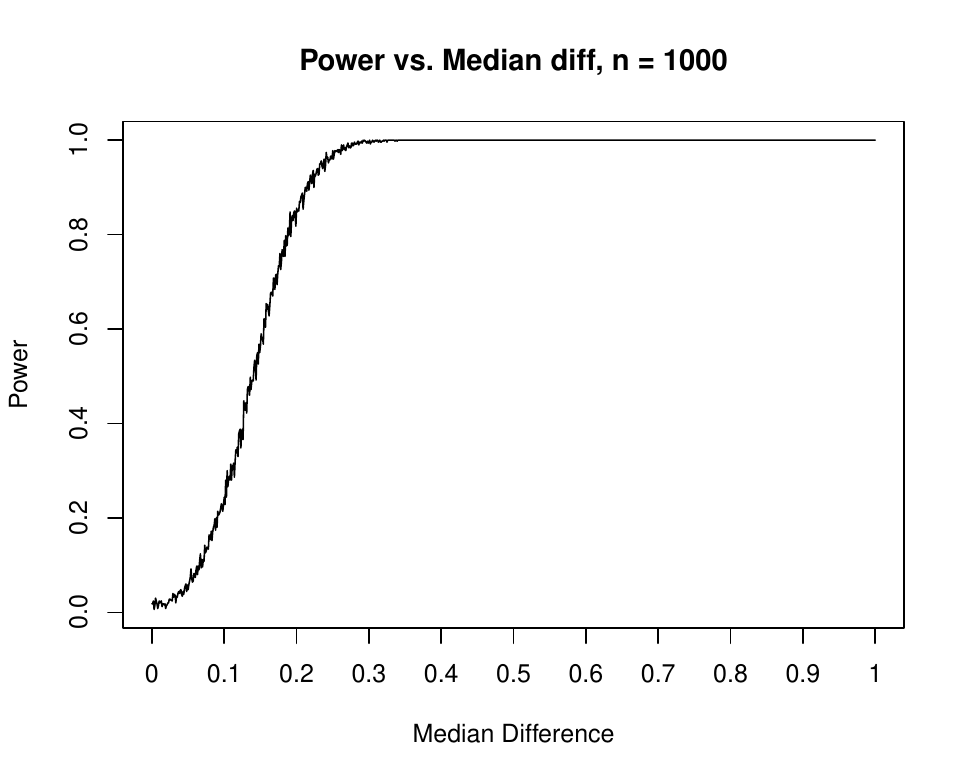}
    \caption{Plot for the power of the test for two samples, each of size 1000 (i.e. $k = 2$ and $n_1 = n_2 = n = 1000$) and population distributions $N(0,1)$ and $N(\Delta,1)$ where $\Delta$ denotes the median difference, plotted in $X-$axis.}
    \label{fig:normal_n1000}
\end{figure}

\begin{figure}[H]
    \centering
    \includegraphics[width=0.6\textwidth]{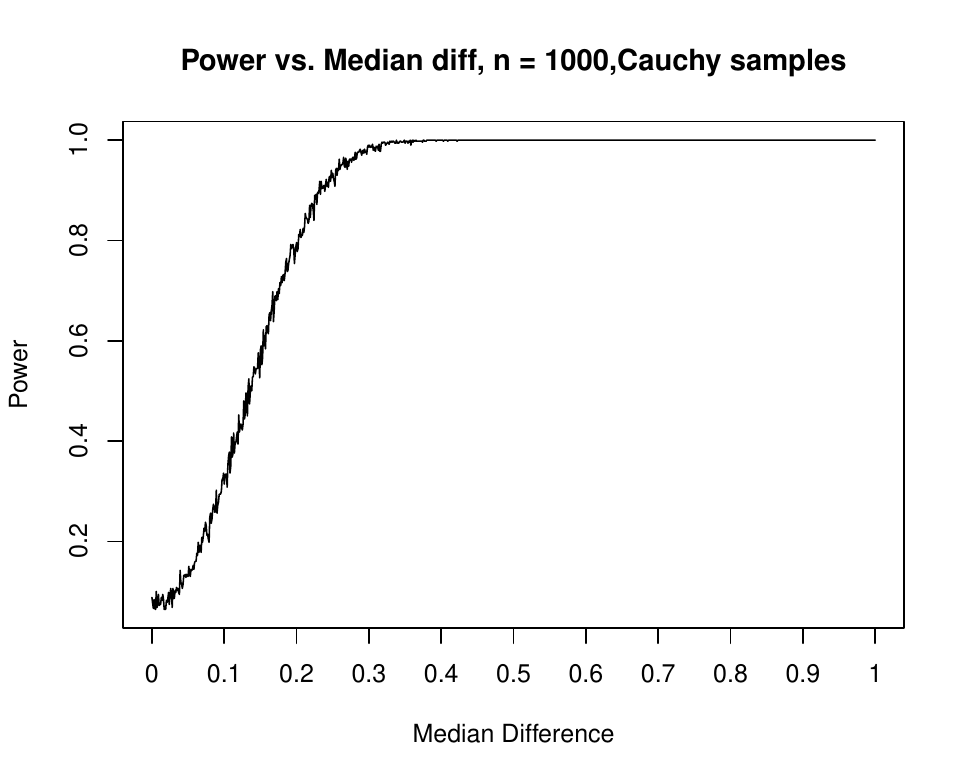}
    \caption{Plot for the power of the test for two samples, each of size 1000 (i.e. $k = 2$ and $n_1 = n_2 = n = 1000$) and population distributions $C(0,1)$ and $C(\Delta,1)$ where $\Delta$ denotes the median difference, plotted in $X-$axis.}
    \label{fig:normal_n1000}
\end{figure}

From the plots, it can be seen that for two samples, if the median difference ($\Delta$) exceeds 0.35 then for sample size 1000, the power of the test is 1. Our test statistic is computationally also very simple. R codes consisting of the function \texttt{median$\_$test} for our testing procedure and codes for reproducing the plots can be find 
\href{https://drive.google.com/file/d/1xhk2uAbdcyO81z-_6d7iQHWF0FCpTnFk/view?usp=sharing}{here}. 

\section{Acknowledgement}
I sincerely thank \textit{Prof. Debashis Paul} (UC Davis and ISI Kolkata) for his valuable feedback and insightful comments on the simulation studies.

\section*{}

\section{Appendix}

\textbf{Proof of Theorem 1:} We begin with the following lemma:

\textbf{Lemma 1:} Let $X_1, \ldots, X_n \overset{\text{iid}}{\sim} F$, having a PDF $f$. Let $\widehat{f}_n$ be the kernel density estimator of $f$ based on $X_1,\cdots,X_n$ such that:
\[
\forall D \subset \mathbb{R}, \; D \text{ compact}, \quad \sup_{x \in D} |\widehat{f}_n(x) - f(x)| \overset{P}{\to} 0.
\]
Furthermore, suppose $f$ is continuous at $M$, where $M$ is the unique population median of $f$. Let $\widehat{M}_n$ be the sample median of $X_1, \ldots, X_n$. Then,
\[
\widehat{f}_n(\widehat{M}_n) \overset{P}{\to} f(M) \quad \text{as } n \to \infty.
\]

\textbf{Proof of Lemma 1:} \[
| \widehat{f}_n(\widehat{M}_n) - f(M) |
\leq | \widehat{f}_n(\widehat{M}_n) - f(\widehat{M}_n) | + | f(\widehat{M}_n) - f(M) | := \Delta_n.
\]
Fix $\varepsilon > 0$. Then since $\widehat{M}_n \xrightarrow{P} M$, 
\[
P\left[\widehat{M}_n \in [M - \varepsilon, M + \varepsilon]\right] = 1 - o(1).
\]
On the event $\widehat{M}_n \in [M - \varepsilon, M + \varepsilon]$, 
\[
| \widehat{f}_n(\widehat{M}_n) - f(\widehat{M}_n) | \leq \sup_{x \in [M - \varepsilon, M + \varepsilon]} | \widehat{f}_n(x) - f(x) | = o_P(1).
\]
Also, $\widehat{M}_n \xrightarrow{P} M$ as $n \to \infty$. Since $f$ is continuous at $M$, 
\[
f(\widehat{M}_n) \xrightarrow{P} f(M) \quad \text{as } n \to \infty.
\]
i.e., 
\[
| f(\widehat{M}_n) - f(M) | = o_P(1).
\]
Therefore, with probability $1 - o(1)$,
\[
| \widehat{f}_n(\widehat{M}_n) - f(M) | \leq \Delta_n,
\]
where $\Delta_n = o_P(1)$. Hence, 
\[
| \widehat{f}_n(\widehat{M}_n) - f(M) | \xrightarrow{P} 0 \quad \text{as } n \to \infty.
\]
Hence the proof of Lemma 1 follows. \hfill $\square$

We now prove Theorem 1. Observe that for each sample, using the asymptotic normality of the sample median, \Cref{ass: sample size} and Slutsk's theorem, we have as $n_i \to \infty$,
\[
\sqrt{n_1} \left( \widehat{M}_i - M_i \right) \xrightarrow{d} N\left( 0, \frac{1}{4 \lambda_i f_i^2(M_i)} \right).
\]
Since all of the samples are assumed to be independent, 
\[
\sqrt{n_1} 
\begin{bmatrix}
\widehat{M}_1 - M_1 \\
\widehat{M}_2 - M_2 \\
\vdots \\
\widehat{M}_k - M_k
\end{bmatrix}
\xrightarrow{d} N(0, \Sigma),
\]
where $\Sigma = \text{diag} \left( \frac{1}{4 \lambda_i f_i^2(M_i)} : 1 \leq i \leq k \right)$. Multiplying with $A$ (defined in \ref{eq: A}), we have 
\[
\sqrt{n_1} A
\begin{bmatrix}
\widehat{M}_1 - M_1 \\
\widehat{M}_2 - M_2 \\
\vdots \\
\widehat{M}_k - M_k
\end{bmatrix}
\xrightarrow{d} N(0, A \Sigma A^\top),
\]
Hence, 
\[
\sqrt{n_1}
\begin{bmatrix}
\widehat{M}_1 - \widehat{M}_2 - (M_1 - M_2)\\
\widehat{M}_2 - \widehat{M}_3 - (M_2 - M_3)\\
\vdots \\
\widehat{M}_{k-1} - \widehat{M}_k - (M_{k-1} - M_k)
\end{bmatrix}
\xrightarrow{d} N(0, A \Sigma A^\top).
\]
Observe that, in view of Theorem 3.1 of \cite{tran1992kernel}, the assumptions of Lemma 1 are satisfied. Thus by \textbf{Lemma 1}, as $n_i \to \infty$, we have $\widehat{\bm{\Sigma}} \convP \Sigma$. Also, $A \widehat{\bm{\Sigma}}A^T$ is almost surely positive definite, hence invertible. Therefore, by the Continuous Mapping theorem and Slutsky's theorem, it follows that 
\[
n_1 \begin{bmatrix}
\widehat{M_1} - \widehat{M_2} - (M_1 - M_2) \\
\widehat{M}_2 - \widehat{M}_3 - (M_2 - M_3) \\
\vdots \\
\widehat{M_{k-1}} - \widehat{M_k} - (M_{k-1} - M_k)
\end{bmatrix}^T
\left( A \widehat{\bm{\Sigma}} A^T \right)^{-1}
\begin{bmatrix}
\widehat{M_1} - \widehat{M_2} - (M_1 - M_2) \\
\widehat{M}_2 - \widehat{M}_3 - (M_2 - M_3) \\
\vdots \\
\widehat{M_{k-1}} - \widehat{M_k} - (M_{k-1} - M_k)
\end{bmatrix}
\xrightarrow{d} \chi^2_{k-1} 
\]
as  $n_1, \ldots, n_k \to \infty$.
Thus the proof of \textbf{Theorem 1} follows. \hfill $\blacksquare$

\textbf{Proof of Theorem 2:} We have, 
\[
M := 
\begin{bmatrix}
M_1 - M_2 \\
\vdots \\
M_{k-1} - M_k
\end{bmatrix}, \quad
\widehat{M} := 
\begin{bmatrix}
\widehat{M}_1 - \widehat{M}_2 \\
\vdots \\
\widehat{M}_{k-1} - \widehat{M}_k
\end{bmatrix}.
\]

Then
\begin{align*}
    & \hspace{0.5cm}\left| \widehat{M}^\top (A \widehat{\Sigma} A^\top)^{-1} \widehat{M} - M^\top (A \Sigma A^\top)^{-1} M \right| \\
    &\leqslant  
\left| \widehat{M}^\top (A \widehat{\Sigma} A^\top)^{-1} \widehat{M} - M^\top (A \widehat{\Sigma} A^\top)^{-1} M \right| +
\left| M^\top (A \widehat{\Sigma} A^\top)^{-1} M - M^\top (A \Sigma A^\top)^{-1} M \right| \\
&= \Delta_1 + \Delta_2 \quad \text{(say)}
\end{align*}
Where
\begin{equation}
    \begin{split}
        \Delta_1 := \left| \widehat{M}^\top (A \widehat{\Sigma} A^\top)^{-1} \widehat{M} - M^\top (A \widehat{\Sigma} A^\top)^{-1} M \right|, \\
        \Delta_2 := \left| M^\top (A \widehat{\Sigma} A^\top)^{-1} M - M^\top (A \Sigma A^\top)^{-1} M \right|.
    \end{split}
\end{equation}
Observe,
\begin{align*}
    \Delta_1 &= \left| (\widehat{M} + M)^\top (A \widehat{\Sigma} A^\top)^{-1} (\widehat{M} - M) \right| \\
    & \leqslant c_1 \|\widehat{M} - M\|_1, \quad \text{with probability } 1 - o(1)
\end{align*}
Now, by \cite{bahadur1966note},
\[
\widehat{M}_i = M_i +  \frac{\frac{1}{2} - \widehat{F}_{n_i,i}(M_i)}{f_i(M_i)} + R_{n_i},
\]
where with probability 1,
\[
R_{n_i} = \mathcal{O}\left(n_i^{-3/4} (\log n_i)^{1/2} (\log \log n_i)^{1/4}\right).
\]

Now,
\[
\widehat{M} - M = 
\begin{bmatrix}
(\widehat{M}_1 - M_1) - (\widehat{M}_2 - M_2) \\
\vdots \\
(\widehat{M}_{k-1} - M_{k-1}) - (\widehat{M}_k - M_k)
\end{bmatrix}.
\]

Hence,
\[
\|\widehat{M} - M\|_1 = R_n + \sum_{i=1}^{k-1} \left| 
 \frac{\frac{1}{2} - \widehat{F}_{n_i,i}(M_i)}{f_i(M_i)} -  \frac{\frac{1}{2} - \widehat{F}_{n_{i+1},i+1}(M_{i+1})}{f_{i+1}(M_{i+1})}
\right|,
\]
where $n = \max\{n_i\}_{i=1}^k$. So,
\begin{equation}
    \Delta_1 \leq c_1 \left[ R_n + \sum_{i=1}^{k-1} \left| 
\frac{1}{2} \frac{\widehat{F}_{n_i}(M_i)}{f_i(M_i)} - \frac{1}{2} \frac{\widehat{F}_{n_{i+1}}(M_{i+1})}{f_{i+1}(M_{i+1})}
\right| \right].
\end{equation}
For the other term, note that 
\begin{align*}
   \Delta_2 &= \left| M^T \left(A  \widehat{\Sigma}A^T\right)^{-1} - \left(A \widehat{\Sigma} A^T\right)^{-1} M \right| \\
   &\leqslant c_2 \left\| \left(A \widehat{\Sigma}A^T\right)^{-1} - \left(A \widehat{\Sigma} A^T\right)^{-1} \right\|_F \\
   &= c_2 \left\| g(\widehat{\Sigma}) - g(\Sigma) \right\|_F
\end{align*}
where $g(\Sigma) := \left(A \Sigma A^T\right)^{-1}$ and $\norm{\cdot}_F$ denotes the Frobenius Norm of Matrices. Now, by Corollary 2.12 of \cite{mcmath1981mean},
\[
\left\| g(\widehat{\Sigma}) - g(\Sigma) \right\|_F \leq \|\widehat{\Sigma} - \Sigma\|_F \sup_{t \in [0,1]} \left\| g'(Z_t(\Sigma, \widehat{\Sigma})) \right\|_F,
\]
where 
\[
Z_t(\Sigma, \widehat{\Sigma}) = t \Sigma + (1-t) \widehat{\Sigma}.
\]
Now, 
\begin{align*}
   g'(\Sigma) &= \frac{\partial}{\partial \Sigma} \left(A \Sigma A^T\right)^{-1} \\
   &= \frac{\partial}{\partial \left(A \Sigma A^T\right)} \left(A \Sigma A^T\right)^{-1} \cdot \frac{\partial}{\partial \Sigma} \left(A \Sigma A^T\right) \\
   &= - \left[ \left(A \Sigma A^T\right)^{-1} \otimes \left(A \Sigma A^T\right)^{-1} \right] \cdot \left[ A \otimes A \right]
\end{align*}
\[
\left[ 
\frac{\partial}{\partial A} A^{-1} = -A^{-1} \otimes A^{-1}, \quad \text{and} \quad
\frac{\partial}{\partial X} (A X B) = B^T \otimes A 
\right].
\]
Therefore,
\[
\sup_{t \in [0,1]} \|g'(Z_t(\Sigma, \widehat{\Sigma}))\|_F = \mathcal{O}(1)
\]
with probability $1 - o(1)$.

Therefore,
\begin{align*}
    \Delta_2 &\leqslant c_2' \|\widehat{\Sigma} - \Sigma\|_F \quad \text{w.p. } 1 - o(1) \\
    &\leqslant c \|\widehat{\Sigma}^{-1} - \Sigma^{-1}\|_F \quad \text{w.p. } 1 - o(1)
\end{align*}
Therefore, we have 
\begin{align*}
    &\hspace{0.5cm}\left| \widehat{M}^T \left(A \widehat{\Sigma} A^T\right)^{-1} \widehat{M} - M^T \left(A \Sigma A^T\right)^{-1} M \right| \\
    &\leqslant \Delta_1 + \Delta_2 \\
    &\leqslant c \left[ R_n + \sum_{i=1}^{k-1} \left|  \frac{\frac{1}{2} - \widehat{F}_{n_i,i}(M_i)}{f_i(M_i)} -  \frac{\frac{1}{2} - \widehat{F}_{n_{i+1},i+1}(M_{i+1})}{f_{i+1}(M_{i+1})} \right| + \|\widehat{\Sigma}^{-1} - \Sigma^{-1}\|_F \right]
\end{align*}
with probability $1 - o(1)$. Hence the proof of \textbf{Theorem 2} follows. \hfill $\blacksquare$
\end{document}